\def\g{$\gamma$}

\def\Msun{\hbox{$\rm\thinspace M_{\odot}$}}

\documentstyle[epsfig]{aipproc}

\begin{document}
\title{Galactic Black Hole Binaries:  High-Energy Radiation}

\author{J.E. Grove$^1$, J.E. Grindlay$^2$, B.A. Harmon$^3$,
X.-M. Hua$^4$, D. Kazanas$^4$, and M. McConnell$^5$}
\address{$^1$E.O. Hulburt Center for Space Research, Naval Research Lab,
Washington, DC 20375\\
$^2$Harvard-Smithsonian Center for Astrophysics, Cambridge, MA 02138\\
$^3$NASA/Marshall Space Flight Center, Huntsville, AL 35812\\
$^4$NASA/Goddard Space Flight Center, Greenbelt, MD 20771\\
$^5$University of New Hampshire, Durham, NH}

\maketitle

\begin{abstract}
Observations of galactic black hole candidates made by the instruments
aboard the
Compton GRO in the hard X-ray and $\gamma$-ray bands have
significantly enhanced our knowledge of the 
emission properties of these objects.
Understanding these observations presents a
formidable challenge to theoretical models of the accretion flow onto the
compact object and of the physical mechanisms that generate high-energy
radiation.  Here we summarize the current state of observations
and theoretical interpretation of the emission from
black hole candidates above 20 keV.

The all-sky monitoring capability of BATSE allows, for the
first time, nearly continuous studies of the high-energy emission from
more than a dozen black hole candidates.  These long-term datasets are
particularly well-suited to multiwavelength comparison studies, from
the radio upward in frequency (Zhang et al. 1997a, these proceedings).
Energy spectral evolution and/or spectral state
transitions have been observed from many of the black hole candidates.
Moderately deep searches of the galactic plane suggest a deficit of weak
$\gamma$-ray transients.  Such population studies have implications for
the origin of black hole binaries and the nature of accretion events.

Observations above 50 keV from OSSE demonstrate that
in the $\gamma$-ray band there exist two spectral states that appear to
be the extensions of the X-ray low (hard) and high (soft), or
perhaps very high, states.
The former state, the ``breaking'' state, 
cuts off with e-folding energy $\sim$100 keV and has
its peak luminosity near this energy; thus substantial corrections need
to be made to historical estimates of the bolometric luminosity of
black holes in the ``low'' state.  In contrast, in the X-ray high (soft)
state, the luminosity peaks in the soft X-rays and the spectrum extends with an
unbroken power law, even up to energies above 500 keV in some cases.
COMPTEL has detected emission above 750 keV from Cyg X-1 
and the transient GRO~J0422+32.  In both cases
the data suggest that an additional weak, hard spectral component is required
beyond that observed by OSSE at lower energies, although
the precise spectral form is yet to be determined.

The breaking $\gamma$-ray
spectrum can be well modeled by Comptonization of soft photons from the
accretion disk in a hot thermal plasma. However, recent studies of the combined
X-ray and $\gamma$-ray spectrum of Cyg~X-1 
and GX339--4 cast severe doubts on the simple geometry of a
hot corona overlying a thermal accretion disk. Furthermore, timing studies of
the former source are inconsistent with spectral formation by Compton scattering
in a uniform, compact hot cloud, suggesting instead a decline in electron 
density with increasing radius.  The power-law 
$\gamma$-ray spectral state creates
more significant theoretical challenges, particularly in explaining the
lack of a break at energies exceeding the electron rest mass.
It has been
suggested that in the X-ray high (soft) state, the high-energy emission
arises from bulk-motion Comptonization in the
convergent accretion flow from the inner edge of the accretion disk.  Such
a process can conceivably generate the $\gamma$ ray spectrum extending without
a cutoff, if the accretion rate approaches that of Eddington.
\end{abstract}

\section{Introduction}

The most reliable evidence for the presence of a black hole in a binary
system comes from determination of a mass function 
through optical measurements of
the radial velocity of the companion star.  If the resulting lower limit
on the mass of the compact object exceeds 3$\Msun$, the upper limit for
the mass of a stable neutron star based on current theory, then
one can reasonably assume that the compact object is a black hole.  There
are at least nine X-ray binary systems with minimum mass estimates exceeding
3$\Msun$, of which three (Cyg~X-1, GRO~J0422+32, and GRO~J1655--40) have
been clearly detected by GRO instruments.  Other objects are identified
as BHCs based on the similarity of their high-energy spectra and 
rapid time variability to those of Cyg~X-1.  Such classification is, of course,
somewhat tenuous.  Before neutron stars and black holes can be reliably
distinguished based on their X-ray and $\gamma$-ray spectra, the full
range of spectral forms from both classes must be observed and characterized.
Extensive knowledge of the X-ray emission of these objects has accumulated
in the literature, but the broad nature of the $\gamma$-ray emission is only
now coming to light, with the high sensitivity of current-generation
instruments.

The instruments of the Compton GRO have made extensive observations 
in the hard X-ray and $\gamma$-ray bands of galactic black hole candidates
(BHCs).  With its all-sky capability, BATSE has monitored emission 
on a nearly continuous basis from
at least three persistent sources (Cyg~X-1, 1E1740.7--2942, GRS~1758--258)
and eight transients
(GRO~J0422+32, GX339--4, N Mus 1991, GRS~1716--249, GRS~1009--45, 4U~1543--47, 
GRO~J1655--40, and GRS~1915+105).  Lightcurves are presented below in
Fig. \ref{lightcurve}.  OSSE has made higher-sensitivity, 
pointed observations of all of these
sources, spectra of which appear below in Fig. \ref{two_state}.
COMPTEL has detected emission above 750 keV from Cyg~X-1
and GRO~J0422+32.  To date, there have been no reported detections 
of galactic BHCs by EGRET.

The French coded-aperture telescope Sigma on the Russian Granat
spacecraft has imaged at least a dozen BHCs, including most of those
in the list above, but with the addition of TrA~X-1, 
GRS~1730--312, and GRS~1739--278.  The latter two objects were weak
transients discovered during a multi-year survey of the galactic center
region and have been
classified as BHCs by their outburst lightcurves and the hardness of 
their spectra (Vargas et al. 1997).  In this survey, Sigma regularly 
detected the
persistent, variable sources 1E1740.7--2942 and GRS~1758--258, both of
which are classified as BHCs on spectral grounds.  The most striking
result from Sigma observations of BHCs is the detection of broad
spectral features below 500 keV from 1E1740.7--2942 (Sunyaev et al. 1991,
Bouchet et al. 1991, Churazov et al. 1993, Cordier et al. 1993)
and N Mus 1991 (Goldwurm et al. 1992, Sunyaev et al. 1992).  These
features have been interpreted as thermally broadened and red-shifted
annihilation radiation from the vicinity of the compact object.

\section{BATSE Survey for Black Hole Binaries in the Galactic Plane}

BATSE has proven to be a remarkably effective tool for the study 
of persistent and transient hard X-ray sources using the occultation 
technique. Not only can known sources be studied, with sensitivity 
of $\sim$100 mCrab for a 1-day integration (Harmon et al. 1992), but 
the powerful occultation imaging technique (Zhang et al. 1993) has 
opened the way for the study of relatively crowded fields and previously 
unknown sources.  Grindlay and coworkers have begun a survey of the
galactic plane with the objective of measuring or constraining
the black hole X-ray binary (BHXB)
population in the Galaxy. BHXBs, both those with 
low mass companions (e.g. the X-ray novae 
such as N Muscae 1991) and the high mass systems such 
as Cyg X-1, are distinguished by having relatively luminous hard X-ray 
(20--100 keV) emission as compared to the systems containing 
neutron stars (Barret et al. 1996a). Thus a survey for 
persistent or transient sources in the hard X-ray band is optimally suited 
for the detection and study of BHXB systems.  Furthermore, since most 
BHXBs are now known to be transient---and indeed the X-ray novae 
have allowed the most convincing dynamical mass function measurements 
of the optical counterparts when they are in quiescence (cf. van Paradijs 
and McClintock 1995)---the transients are the systems most likely to be BHXBs 
(and see van Paradijs 1996 for a likely explanation). 

Given the recurrence times of the prototypical soft X-ray transients 
(SXTs) to be $\gtrsim$10-50 years (e.g. $\gtrsim$10 years for Nova
Muscae; cf. Barranco and Grindlay 1997), and the 
relatively nearby (1--4 kpc) optical distances for
the bright SXTs identified at a rate of $\sim$1 per year by Ginga, WATCH
and BATSE, it is straightforward to extrapolate 
that 3--4 per year should be detectable by BATSE in 
these deeper searches (i.e. with peak fluxes at or below 100 mCrab)
if the BHXBs are distributed uniformly in a galactic disk 
with radius 12 kpc.  Thus over the 6 years of archival BATSE data now 
available, a full analysis of
the galactic plane should yield a significant number ($\gtrsim$20) of new
BHXBs. 

The CfA BATSE Image Search (CBIS) system (Barret et al. 1997, Grindlay
et al. 1997) has been run on 900d of BATSE data.  
Known sources, though perhaps not previously detected by BATSE 
(e.g. Cir X-1; Grindlay et al. 1997) are 
found in the survey.  Both short-outburst BHXBs (e.g. 4U1543--47) and longer 
duration outburst from neutron star LMXBs
(e.g. 4U1608--52) have been found in the data ``automatically'' at the 
times and approximate fluxes seen with direct occultation light curve 
analysis (Harmon et al. 1992).  More importantly, the search has yielded
at most 5 candidate new (i.e. uncatalogued) sources.
All of these candidate sources are below
$\sim$50--100 mCrab in peak flux; no new transients at $\sim$100--200
mCrab have been found in this initial survey.  The preliminary results  
thus suggest, but do not yet prove, 
a lower rate of candidate new BHXB transients than the simple
scalings above would suggest:  a rate of $\sim$1--2 per year rather than the 
3--4 possibly expected.

If the true number is in fact much less than expected,
then important questions arise
for the BHXB population and/or transient outburst models: 

\noindent{\bf 1.}{\it What is the total number of BHXBs?} Many authors
(e.g. Tanaka and Lewin 1995 and references therein) estimate on the
basis of simple arguments (such as the SXT detection rate) that the total
population of low-mass BHXBs in the galaxy is $\gtrsim$300--1000. If 
BATSE cannot
find the predicted fainter systems, have the optical distances been 
systematically under-estimated or are the peak luminosities typically
fainter?

\noindent{\bf 2.}{\it What is the formation rate of BHXBs?} Clearly,
a measurement of, or constraints on, the total number of BHXBs is needed to 
determine if models for the formation of BHXBs vs. neutron star binaries 
(e.g. Romani 1992) that predict large
numbers of low-mass BHXBs are correct.

\noindent{\bf 3.}{\it What are the characteristic spectra and 
lightcurves of SXTs?} A search for 
faint BHXBs with BATSE (and future more sensitive
surveys) is crucial for comparison with ASM or WFC searches (RXTE and future
ASMs) which will typically operate in the 2--10 keV band. Recent WFC 
detections of $\sim$10 mCrab transients in the galactic bulge region by
SAX (e.g. in 't Zand and Heise 1997) may suggest a much
higher rate of either fainter or softer transients.

\noindent{\bf 4.}{\it What are the characteristic recurrence times for SXTs?} 
If the BATSE detection rate is much lower than predicted, is the
recurrence time much longer than the small-N statistics would now
indicate?  This of course directly affects questions 1 and 2 above as
well as the outburst models. It is of particular interest to search
for ``mini''-outbursts of the nearby bright systems (e.g. A0620--00),
which might be expected to be more detectable in the hard X-ray band
than the soft band in the low (hard) state (cf. Fig. \ref{two_state}).

\section{Two gamma-ray spectral states}

The historical record of X-ray (i.e. $<$30 keV) observations
of galactic BHCs reveals at least four spectral states
(see, e.g., Tanaka 1989 \& 1997,
Grebenev et al. 1993, and van der Klis 1994 \& 1995).
In the ``{\em X-ray very high}'' and
``{\em X-ray high (soft)}'' states, the X-ray spectrum is dominated
by an ``ultrasoft'' thermal or multi-color blackbody 
component with kT $\sim$ 1 keV.  A weak power-law tail, with photon
number index $\Gamma \sim $ 2--3, is frequently present and dominant
above $\sim$10 keV.  The states differ in X-ray luminosity---the former
being close to the Eddington limit, $L_E$, and the latter typically
$\sim$3--30 times less than $L_E$---and in the character of their
rapid time variability---the very high state usually has 
3--10Hz quasi-periodic oscillations (QPOs) and 
stronger broad-band noise. 
The ``{\em X-ray low (hard)}'' state exhibits a single
power-law spectrum with $\Gamma \sim $ 1.5--2, and
a typical X-ray luminosity of $<$1\% of Eddington.  Recent measurements
by OSSE indicate that, in this state,
the \g-ray luminosity of a typical BHC exceeds the X-ray by
a factor $\sim$5 (Grove et al. 1997b).  This state is characterized
by strong, rapid, intensity variability, with rms variations of order
a few tens of percent of the total emission.  The ``{\em X-ray off}'' or
``{\em quiescent}'' state exhibits very low level emission with
uncertain spectral shape at a luminosity $L_X < 10^{-4} L_E$.

\begin{figure}
\centerline{\epsfig{file=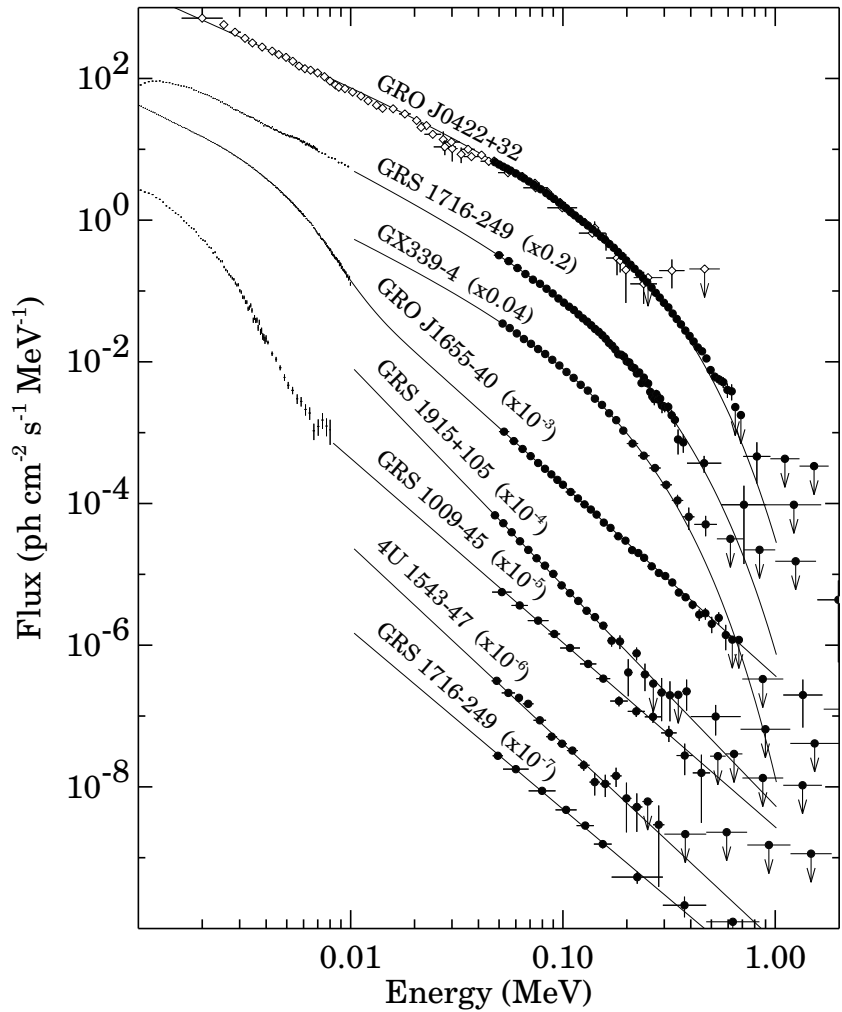,width=12.2cm}}
\caption{Photon number spectra from OSSE for seven transient 
BHCs.  Spectra are averaged over all observing days for which there
was detectable emission
and, for clarity of the figure, have been scaled by arbitrary factors
as indicated.  Two spectral states are apparent.
Contemporaneous TTM and HEXE data (open diamonds) and ASCA data (crosses)
are shown for GRO~J0422+32 and GRS~1716--249, respectively.
Non-contemporaneous ASCA data (crosses)
are shown for GRS~1009--45 and GRO~J1655--40.  ASCA data for 
GRS~1716--249 and GRS~1009--45 are from Moss (1997).  Figure is
from Grove et al. (1997b).}
\label{two_state}
\end{figure}

OSSE observations of a number of transient
BHCs (Grove et al. 1997a, 1997b) indicate that there
are at least two distinct $\gamma$-ray spectral states, the
``{\em breaking}'' state, which corresponds to the X-ray low (hard) state,
and the
``{\em power law}'' state, which corresponds to the X-ray high (soft) state 
or very high state.  The identification of the breaking state with the
low (hard) state is quite firm but, because of the paucity of simultaneous
X-ray and $\gamma$-ray observations, the identification of the power law
state with which of the two high states is less certain.  It is clear,
though, that the presence of a strong soft excess at least some of the time
during an outburst is a requirement for the $\gamma$-ray power law state
to be observed.
Fig. \ref{two_state} shows photon number spectra from OSSE for seven
transient BHCs, along with the best-fit analytic model extrapolated down
to 10 keV and contemporaneous X-ray data if they are available.  The two
spectral states are readily apparent.  

Sources in the X-ray low (hard) [i.e. breaking $\gamma$-ray] 
state typically have
an X-ray index $\Gamma \sim 1.5$, begin breaking from
the power law at $E_b \sim 50$ keV, and cut off with exponential folding
energy of $E_f \sim 100$ keV.  The bulk of the luminosity in this state
is emitted near 100 keV.  Indeed the spectra of GRO~J0422+32 and
GRS~1716--249 indicate that the luminosity above 50 keV (or 10 keV) exceeds
that in the 0.5--10 keV band by a factor of $\simeq$4 (or $\simeq$6).

Sources in the \g-ray power-law state have a strong ultrasoft excess 
above a single power-law spectrum,
with $\Gamma \sim 2.5-3.0$ and no evidence for a high-energy break, even
at energies exceeding $m_e c^2$.  No
spectral features (e.g. narrow or broad lines) are apparent near or
above 511 keV, as would be expected from standard nonthermal
Comptonization models (Blumenthal \& Gould 1970,
Lightman \& Zdziarski 1987).  In this state, $L_\gamma$ is only
a fraction of the X-ray luminosity.  The strong ultrasoft component
is the signature of either the X-ray very high or high (soft) state,
although the generally weak or absent {\em rapid} time-variability ($<$1 sec)
suggests that the association is with the high (soft) state
(Grove et al. 1997b).
As a caveat, we note that there indeed can be a degree of independence of the
ultrasoft and power-law spectral components exhibited in the 
long-term temporal behavior
of the broadband emission:  for example, while recent RXTE observations 
(Greiner, Morgan, \& Remillard 1996) of
the galactic superluminal source GRS~1915+105 demonstrate that the
X-ray intensity is dramatically variable on timescales of tens to
thousands of seconds, showing several repeating temporal structures,
the $\gamma$-ray emission is steady and only slowly evolving
(Grove et al. 1997b).

Recent observations of Cyg~X-1 reveal
a bimodal spectral behavior in the $\gamma$-ray
band as well as the X-ray band (Phlips et al. 1996),
equivalent to the above, and confirming the identification of X-ray
and $\gamma$-ray states (Gierlinski et al. 1997; Phlips et al. 1997).
These observations are discussed in greater detail below.

The transient GRS~1716--249 is shown twice in Fig. \ref{two_state},
apparently having undergone a spectral state change late
in its second outburst (Moss 1997, Grove et al. 1997b).  As is the case for
Cyg~X-1 (Phlips et al. 1996), the power-law state has lower 
$L_\gamma$ than the breaking state.  How the
{\em bolometric} luminosity changes is less certain and will require further
simultaneous broadband observations.

\begin{figure}
\centerline{\epsfig{file=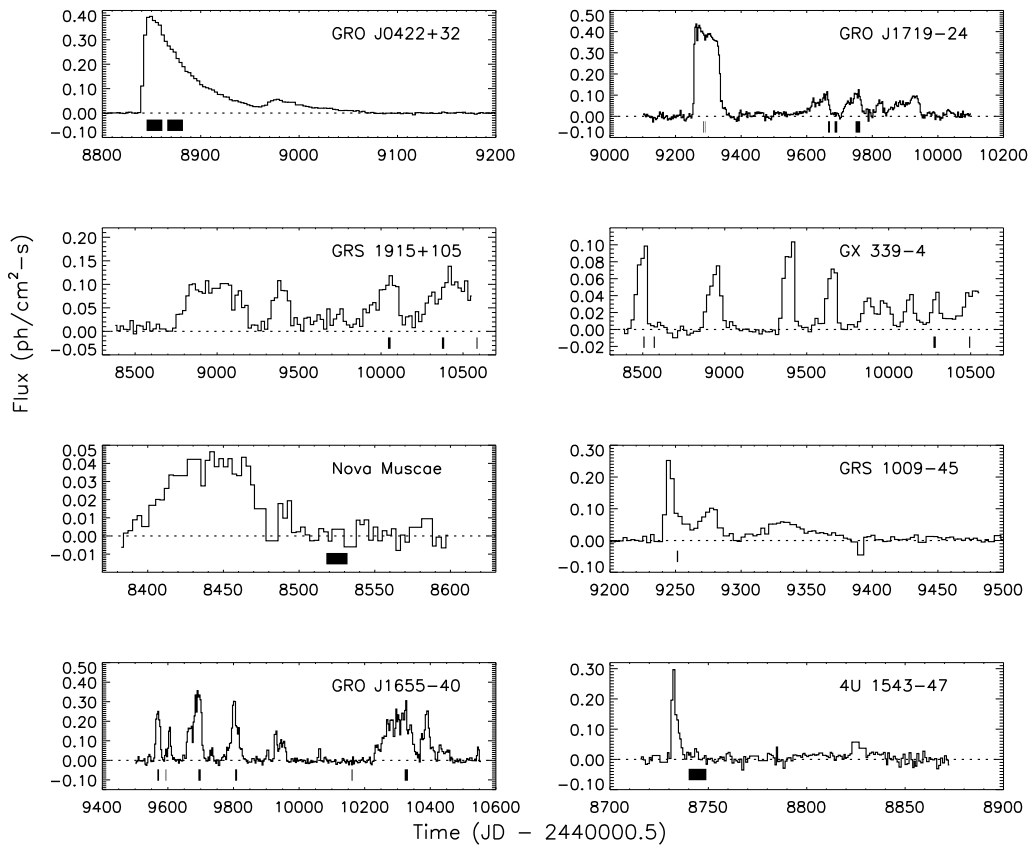,width=13.5cm}}
\caption{
Lightcurves of transient black hole binary systems in the 20--100 keV
band detected with BATSE.  OSSE observing times are shown as bold solid lines 
along the horizontal (time) axis.  All light curves include the primary
and secondary outbursts until mid-1997, except for N Mus 1991.  Its 
primary outburst occurred prior to launch of CGRO.}
\label{lightcurve}
\end{figure}

BATSE hard X-ray (20--100 keV) lightcurves for the same seven 
transient BHCs plus N Mus 1991 are shown in Fig. \ref{lightcurve}.  
Harmon et al. (1994) identified two types of transients based on
such lightcurves.  The first type has a relatively 
fast rise followed by a more gradual,
decaying flux envelope after the initial outburst, and secondary maxima 
a few weeks to months
into the decline.  The decay sometimes approaches an exponential form
as seen in the soft X-ray band (Tanaka and Lewin 1995).  The rise and
decay times vary over a broad range, but are order of a few days and 
few tens of days, respectively.  The second type has a longer rise time
(of order weeks) and multiple, recurrent outbursts of highly variable
duration.  Major outbursts tend to be of similar
intensity, with active periods lasting for several years.   GRO J0422+32, 
N Mus 1991, GRS~1009--45 and 4U~1543--47 
are of the first type, and GRS~1915+105 and GX339--4 are of the second
type.  GRS~1716--249 ($=$GRO~J1719--24) and GRO~J1655--40 seem to share
properties of both types.  No coherent, long-term periodicities in the
high energy emission have been seen in either type of 
transient.  There appears to be no correlation between the lightcurve
type and the spectral state.


\begin{figure}
\centerline{\epsfig{file=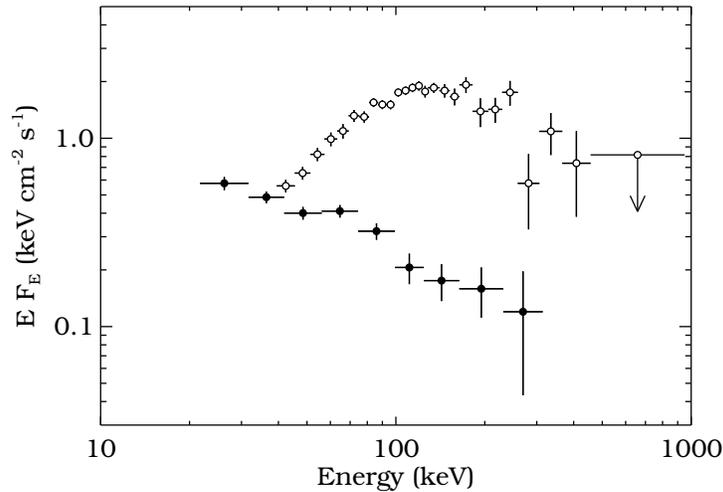,width=4.0in}}
\vspace{10pt}
\caption{Spectra of 1E1740.7-2942 in the $\gamma$-ray breaking state
(open circles; OSSE) and power-law state (solid circles; BATSE).  The
OSSE data have been corrected for nearby point sources and diffuse 
emission.}
\label{1e_spec}
\end{figure}

Elsewhere in this volume, Zhang et al. (1997a) report the detection of
both $\gamma$-ray spectral states from 1E1740.7--2942, the
persistent BHC near the galactic center.  The breaking
$\gamma$-ray state has higher $L_\gamma$ than the power law state.  They also
report the detection of two luminosity states in GRS~1758--258, but
because of the relative faintness of the source, do not present any
spectral analysis.  In Fig. \ref{1e_spec} we have plotted the 
spectrum of 1E1740.7--2942 in the low-luminosity, power-law state
from BATSE (adapted from Zhang et al.) and the high-luminosity,
breaking state from OSSE.  Because neither BATSE nor OSSE is an imaging
instrument, both measurements are subject to contamination from the
bright diffuse galactic continuum emission and nearby point sources.
To minimize this potential source of error, the BATSE data were 
collected when the limb of the earth was at large angles to 
the galactic plane, so that the galactic ridge component would not produce any 
occultation modulation (S.N. Zhang, private communication).  The OSSE
spectrum shown is the result of a simultaneous fit to several known
point sources, which were reasonably well separated by 
scanning the OSSE detectors
along the ecliptic, and we have subtracted an estimate of the 
diffuse emission derived from
an extensive series of galactic-center region mapping observations
(G.V. Jung, private communication).

\section{Broadband observations and spectral modeling}

In recent years, substantial progress has been made in our understanding of
the environment of accreting black holes and the physical processes that
drive the high-energy emission.  Indeed the breaking $\gamma$-ray
state can be relatively well modeled by Comptonization of soft photons from the
accretion disk in a hot thermal plasma, with the plasma temperature determined
by the cutoff energy.  In some cases, additional spectral components 
are required  which can be understood as effects of the hard radiation
scattering off relatively cold electrons  (for example as a  reflection 
off a cold accretion disk or the transmission of the hard radiation
through a cold medium surrounding the compact source).
The nature and geometry of the Comptonizing
plasma is, however, highly debatable, and recent studies of the combined
X-ray and $\gamma$-ray spectrum and temporal properties of Cyg~X-1 
indicate that a simple, slab geometry for the corona is implausible.  

\begin{figure}
\centerline{\epsfig{file=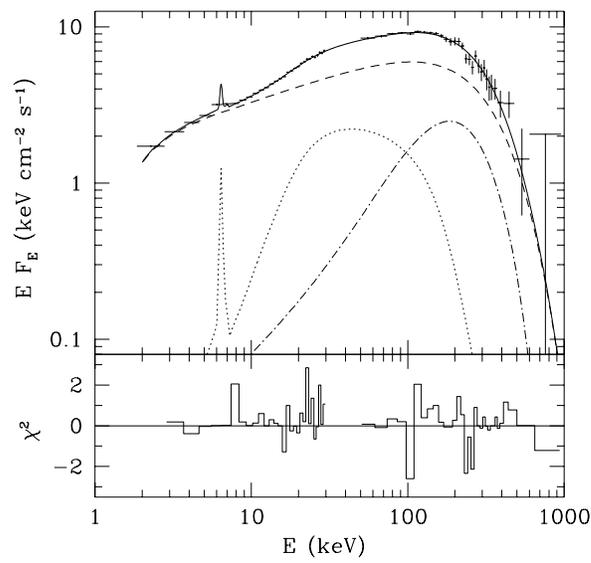,width=3.5in}}
\vspace{10pt}
\caption{Broadband spectrum of Cyg~X-1 observed simultaneously by ASCA
\& OSSE.  Two-temperature thermal Comptonization model (dashed and
dot-dashed curves), along with Compton reflection (dotted curve).
Bottom panel gives contribution to total $\chi^2$.  From
Gierlinski et al. (1997).}
\label{cyg_spec}
\end{figure}

Simultaneous, broadband spectral measurements
are especially powerful in elucidating the physical processes that drive
the high-energy emission of BHCs.  Two recent broadband studies, one from
Ginga and OSSE observations of Cyg~X-1 (Gierlinski et al. 1997) and the
other from ASCA and OSSE observations of GX339--4 (Zdziarski et al. 1997),
are the most detailed investigations to date on BHCs in the X-ray low
(hard) state.

In the low (hard) state, the X-ray spectrum of Cyg~X-1 consists typically
of a power law with photon number index $\Gamma \sim 1.6-1.7$ and a
Compton-reflection component that significantly affects the continuum
above 10 keV and includes near 7 keV both an Fe K edge and an Fe K$\alpha$
fluorescence line (Ebisawa et al. 1996 and references therein).  The
low-energy gamma-ray spectrum is steeply cut off above $\sim$150
keV and, while $L_\gamma$ varies by a factor of at least several in this
state, the spectral parameters vary only weakly (Phlips et al. 1996).
The X-ray (Ueda et al. 1994) and $\gamma$-ray (Grabelsky et al. 1995)
spectra of GX339--4 in the low (hard) state
are generally quite similar in form to those of Cyg~X-1.

For both Cyg~X-1 and GX339--4, the 
combined X-ray and $\gamma$-ray spectrum was modeled by
thermal Comptonization in a spherical cloud with an isotropic source
of soft seed photons from a blackbody distribution, as described in
Zdziarski, Johnson, \& Magdziarz (1996).  The parameters of the hot plasma
are the temperature, $kT$, and the X-ray photon number index, $\Gamma$,
which is related to the Thomson optical depth,
$\tau$, of the Comptonizing plasma.  The model allowed for Compton reflection
with reflector solid angle $\Omega$, and an Fe K$\alpha$ line.
An additional soft X-ray blackbody component was required below 3 keV
from both sources; this component is likely to originate in
a cold accretion disk in the vicinity of the hot Comptonizing plasma.
This same cold disk would then also be responsible for the Compton
reflection component.  In the case of Cyg~X-1, the spectrum cut off
above 100 keV too sharply for an isotropic, single-temperature
Comptonization model, so Gierlinski et al. (1997) added a Wien-like
component from an optically-thick plasma at $kT \sim 50$ keV, speculating
that it is the signature of a transition region between the hot and
cold media in the accretion flow.

\begin{figure}
\centerline{\epsfig{file=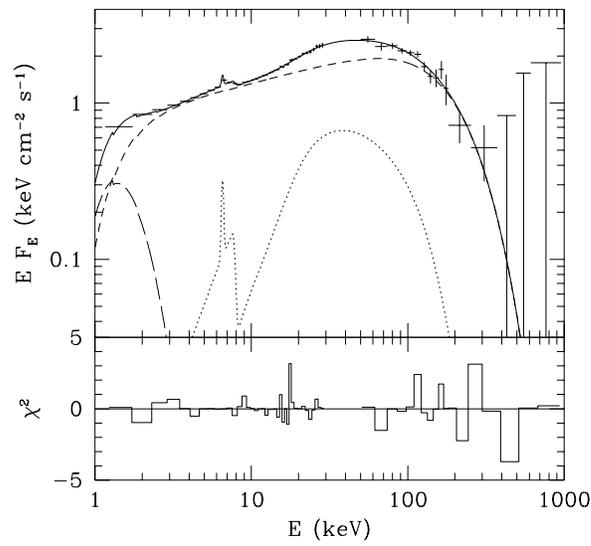,width=3.5in}}
\vspace{10pt}
\caption{Broadband spectrum of GX339--4 observed 
simultaneously by Ginga \& OSSE
on 1991 Sep 11.  Model consists of blackbody radiation (long dashes) as
seed for thermal Componization in hot plasma (short dashes), which is
Compton reflected from the disk (dotted).  Bottom panel gives contribution
to total $\chi^2$.  From Zdziarski et al. (1997).}
\label{gx339_spec}
\end{figure}

As shown in Fig. \ref{cyg_spec} and \ref{gx339_spec}, these complicated,
multi-parameter models give an excellent description of the X-ray and
OSSE data in the entire energy range, from $\sim$2 keV to $\sim$1 MeV.
The time-averaged spectrum of Cyg~X-1 has been measured by COMPTEL
out to several MeV (McConnell et al. 1997).  A comparison with
contemporaneous data from both OSSE and BATSE has so far yielded
only limited insights because of discrepancies among the spectra from
the three instruments.  While the nature of these discrepancies is still
being investigated, they may result merely from a difference in absolute
flux normalization.  In any case, it is becoming clear that the emission
above 1 MeV implies more than a simple Comptonization process.  A similar
conclusion can be drawn from COMPTEL results for GRO~J0422+32
(van Dijk et al. 1995).

For both Cyg~X-1 and GX339--4, the description of Gierlinski et al.
(1997) and Zdziarski et al. (1997) rules out a geometry with a 
corona above the surface of an optically-thick disk (Haardt \& Maraschi 
1993; Haardt et al. 1993).  The solid angle $\Omega$ of the reflector 
is only 30--40\% of the 
$2 \pi$ expected for a corona above a flat disk.  Furthermore, the
intrinsic spectrum is so hard that the flux of the incident seed photons
is about an order of magnitude less than the Comptonized flux
(i.e. the source is ``photon starved''), whereas the seed photon
Comptonized photon fluxes 
are almost the same in the disk-corona model
(Haardt \& Maraschi 1993).  In addition for Cyg~X-1, the
narrowness of the K$\alpha$ line observed by ASCA 
(Ebisawa et al. 1996) implies that the
reflecting medium is cold and therefore likely far from the central
black hole.  On the other hand, the fits {\em can} correspond
to a geometry with a hot inner disk surrounded by a colder outer disk.
The hot disk is geometrically thick and irradiates the outer cold disk.
The irradiation gives rise to the Compton-reflection component with
small $\Omega$, and accounts for the modest blackbody emission of the
disk observed at low energies.

There are currently no high-sensitivity, {\em simultaneous} X-ray 
and \g-ray studies in the literature for the X-ray high (soft) state, 
although several simultaneous
datasets from RXTE and GRO exist for the superluminal sources
GRO~J1655--40 and GRS~1915+105.
A simultaneous ASCA and BATSE observation of GRO~J1655--40 {\em has}
been published (Zhang et al. 1997b), which shows a strong ultrasoft
excess and a soft power-law tail ($\Gamma \simeq 2.4$) extending 
beyond 100 keV.

The physical processes driving the hard emission in
the X-ray high (soft) state are less well determined.  It is generally
well agreed upon that the ultrasoft component, which can be described
by a multi-color disk blackbody spectrum (Mitsuda et al. 1984), is
thermal emission from an optically thick and geometrically thin accretion
disk roughly in the region $10^7 < r < 10^9$ cm from the black
hole.  The power-law tail seen by Ginga from a number of BHCs in
the high (soft) state (e.g. GX339--4:  Makishima et al. 1986) was
ascribed to thermal Comptonization in the hot ($kT \sim 60$ keV) 
inner disk region.  This interpretation is ruled out by Sigma observation
of GX339--4 in this state with a spectrum that extends
unbroken to at least 100 keV (Grebenev et al. 1993), well beyond the break for
a Comptonized spectrum of this temperature.  The unbroken power laws observed
by OSSE from a number of other BHCs require temperatures of at least several
hundred keV.

Ebisawa, Titarchuk, \& Chakrabarti (1996) propose that the high-energy 
emission in the high (soft) state arises from bulk-motion Comptonization of 
soft photons in the convergent accretion flow from the inner edge
of the disk.  In the high (soft) state, the copious soft
photons from the disk cool the electrons efficiently in the inner,
advection-dominated region, and Comptonization due to bulk motion dominates
over that due to thermal motion.  In contrast, in the low (hard) state,
there are fewer soft photons, hence higher temperatures in the
Comptonization region (i.e. $\sim$50--100 keV, rather than $\sim$1 keV), and
thermal Comptonization dominates.
Calculations by Titarchuk, Mastichiadis, \& Kylafis (1996) indicate that
the bulk-Comptonization spectrum can continue unbroken well beyond
$m_e c^2$, as is observed at least in the case of GRO~J1655--40.

The detection of Compton reflection in the high (soft) state may prove
to be a complication (A. Zdziarski, private communication).  Compton 
reflection is clearly 
seen in the low (hard) spectrum of a number of sources, and is detected
in both {\em spectral states} in N Mus 1991 (Ebisawa et al. 1994) and Cyg~X-1
(Gierlinski et al. 1997).  In the latter objects, the reflection 
parameter $\Omega$ approaches
2$\pi$ in the high (soft) state, indicating that the source of the hard
photons completely covers the reflector, e.g. the cool disk.
The bulk-Comptonization model postulates an advecting, Thomson-thick
medium above the disk, and it is in this advecting medium that the
Comptonization takes place.  However, because it is optically thick,
the reflected component is trapped in the advecting flow and does not
escape to the observer.  Thus detection of reflection in the high (soft)
state would cast doubt on the application of bulk Comptonization in
these objects.

\section{The search for line emission}

Transient, broad emission lines have been reported in the literature from 
Cyg~X-1 (the ``MeV bump''), 1E1740.7--2942,
and N Mus 1991.  In the latter two cases,
the lines have been interpreted as red-shifted and split (due to disk
rotation) annihilation features.  The very high sensitivity of the GRO
instruments relative to previous gamma-ray telescopes makes deep searches
for such lines possible on many timescales.  Long-term
observations of Cyg X-1, by both COMPTEL and OSSE, at a range of hard X-ray
intensities set strict upper limits on the magnitude and duty cycle of a
possible MeV bump.  The upper limits are more than an order of magnitude below
the historical reports (Phlips et al. 1996, McConnell et al. 1994).  Similarly, 
searches by Grove et al. (1997b) through hundreds of days
of BHXB observations by OSSE have 
revealed no evidence for transient red-shifted
annihilation lines at levels approaching those in the literature.
For example, a broadened 480 keV line at $6 \times 10^{-3}$ ph cm$^{-2}$
s$^{-1}$, the intensity reported from N Mus 1991 by Sigma 
(Goldwurm et al. 1992, Sunyaev et al. 1992),
would have been detected by OSSE at $\sim$40$\sigma$ in an average
24-hour period.  

In general these transient events are reported by a single instrument and
cannot be confirmed or
refuted because no other instrument is observing at the same
time.  However, the broad-line excess from 1E1740.7--2942 in 1992 Sep
reported by Sigma (Cordier et al. 1993) 
is {\em not} confirmed by OSSE (Jung et al. 1995), which by
chance was viewing the galactic center region before, during, and after
the event.  Confirmation of such transients would provide strong support for
pair plasma models of black hole radiation, lending credence to the
suggestion (e.g. Ramaty et al. 1994)
that sources such as 1E1740.7--2942 might be significant
sources of positrons in the central region of our
galaxy.

\section{Rapid variability and phase lags}

Strong, rapid, aperiodic
variability (i.e. on timescales of tens of seconds or less)
is frequently reported in X-rays from BHCs in the X-ray low (hard) state,
with the occasional appearance of peaked noise or quasi-periodic
oscillations (QPO).  The rms variability is typically of order tens
of percent of the average intensity.  For recent
reviews, see van der Klis (1994 \& 1995).  The GRO instruments find 
similarly strong, rapid
variability in gamma-ray emission from sources in the X-ray low (hard) 
state; see e.g. van der Hooft et al. (1996) for GRS~1716--249 and
Grove et al. (1994) for GRO~J0422+32.

Recent results from RXTE indicate that the
superluminal sources GRO~J1655--40 and
GRS~1915+105 in the X-ray high (soft) state
both show weaker rapid X-ray variability (of order several percent rms) and
QPOs on many timescales, at
frequencies up to 67 Hz in the latter case (Morgan, Remillard, \& Greiner
1997), while such
variability is undetected in the gamma-ray band, where the statistical
limits are $\sim$5\% (Crary et al. 1996, Kroeger et al. 1996).

Using the BATSE instrument, van der Hooft et al. (1996) studied the
evolution of the rapid time variability of GRS~1716--249 (GRO~J1719-24)
throughout an entire $\sim$80-day outburst.  The power density spectrum
showed a strong QPO with a centroid frequency that increased from
$\sim$0.04 Hz at the onset of the outburst to $\sim$0.3 Hz at the end.
Interestingly, they reported that the power spectrum could be
described with a single characteristic profile, the frequency scale of
which stretched proportionally during the outburst, and that the total
power, integrated over a scaled frequency interval, was constant throughout
the outburst.

\begin{figure}
\centerline{\epsfig{file=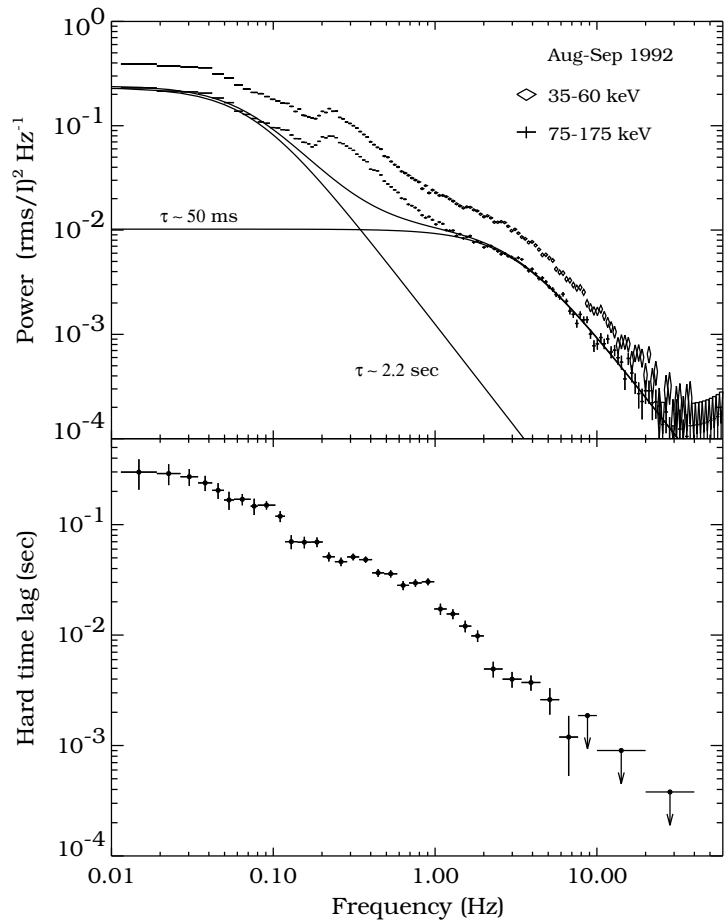,width=4.0in}}
\caption{Upper panel:  OSSE power density spectra of GRO~J0422+32 in
$\sim$35--60 keV and 75--175 keV energy bands.  From Grove et al. (1994).
Lower panel:  Time lag spectrum between same energy bands.}
\label{0422_timing}
\end{figure}

The upper panel of Fig. \ref{0422_timing}, adapted from Grove et al. (1994),
shows the normalized power density spectra (PDS)
in the 35-60 keV and 75-175 keV bands for the OSSE observation of
GRO~J0422+32.  The shape of the power spectrum 
is essentially identical in the two energy
bands.  It shows breaks at a few $10^{-2}$ Hz and a few Hz, and a strong peaked
noise component at 0.23 Hz, with FWHM $\sim0.2$ Hz.  
The peaked noise profile is broad and asymmetric,
with a sharp low-frequency edge and a high-frequency tail; thus the physical
process responsible for the peaked noise appears to have a well-defined maximum
timescale.  We note that, in contrast to GRS~1716--249, the
characteristic frequencies of the shoulders and the peak
are independent of source intensity.
(In passing we also note the striking similarity of the power spectrum of
GRO~J0422+32 to that of the X-ray burster 1E1724--3045 derived from RXTE data,
including both a low-frequency
and a high-frequency break and a strong peaked noise component.  See
Olive et al. in these proceedings for details).

The lower panel shows the time lag spectrum between these
two energy bands:  the hard emission (75--175 keV)
lags the soft emission (35--60 keV) at all Fourier frequencies, falling
crudely as $1/f$, up to about 10
Hz, where there is no statistically significant lag or lead between the two
bands.  In the 10--30 Hz frequency range, time lags as small as 1 ms
would be detectable at $>$99\% confidence.  At frequencies
$\sim$0.01 Hz, hard lags as large as 300 ms are observed.
There is no significant change in the lag at the
frequencies dominated by the strong peaked noise component at 0.23 Hz.

The lag of GRO~J0422+32 as a function of Fourier frequency is quite
similar to the lags reported from Cyg X-1 
(Miyamoto et al. 1988; Cui et al. 1997;
Wilms et al. 1997), N Mus 1991 and GX339--4 (Miyamoto et al. 1993), in 
rather different energy ranges. The OSSE data from GRO~J0422+32 provide
more evidence indicating that the frequency-dependent time lag 
is a common phenomenon shared by many, if not all, accreting objects 
in binaries.

This form of PDS and the
$1/f$ dependence of time lag on the Fourier frequency, first
observed in Cyg X-1 by Ginga (Miyamoto et al. 1988), are very different
from that expected in accretion
models which presumably produce most of the X-ray and $\gamma$-ray
emission from a region whose size is comparable to that of the last stable
orbit around a black hole of mass a few M$_{\odot}$: The characteristic
time scale associated with the dynamics of accretion in such an object
is of order $10^{-3}$ sec, and consequently one would expect most of
the associated power in the kHz frequency range.  By contrast there is a
remarkable {\it lack} of power at this range.

The shape of the PDS of accreting BHC has been the source of much
discussion, but there exists no widely accepted, compelling theory
which provides a ``reasonable" account of it.  While most of the attention
to date has been focused on the $1/f$-like noise in the intermediate
frequency regime ($\sim 0.01 - 10$ Hz), the breaks at both the
low and high frequencies deserve equal attention.  It is puzzling
both that there is a lack of high-frequency power and that most of
the variability power appears
concentrated at the low frequency break, i.e. 4--5 orders of
magnitude lower than the frequencies associated with the dynamics
involved with the production of X-rays.

Generally, the models of BHC variability attempt to reproduce the
observed shapes of the PDS (in fact, mainly the slope of their
power-law section) as an ensemble of {\it exponential} shots with
a range of decay times and/or amplitudes. The ensemble of shots
is usually derived by simulating the accretion onto the BH
in terms of avalanche-type models with Self Organized Criticality
(Bak 1988; Negoro et al 1995).

This approach gives very little physical insight into the breaks
at low and high frequencies, and it may well be erroneous at
the outset.  The reason is shown in Fig. \ref{0422_timing}b, the
time-lag spectrum.  The roughly-$1/f$ dependence of time lag on the 
Fourier frequency is very different from that expected if
the 35--175 keV radiation results from Comptonization of
soft photons by hot electrons in the vicinity of a black hole;
under these conditions the time lags, which are indicative of the
photon  scattering time in the hot electron cloud, should be independent
of the Fourier frequency and of order $10^{-3} $ sec, the photon
scattering time in this region, which is roughly similar to the
dynamical time scale.

The observed lags (Fig. \ref{0422_timing}b) are 
generally much longer and
Fourier-frequency dependent, a fact noticed first by Miyamoto
et al. (1988). The very long observed lags ($\simeq$0.3 sec) and
in particular their dependence on the Fourier frequency exclude
from the outset models in which the variability is produced by
mechanisms which effect a modulation of the accretion rate. These
models, while they are constructed to produce the observed PDS,
since they produce the hard radiation from Comptonization in vicinity of
the black hole, would yield very short, frequency independent
lags.


The discrepancy prompted  Kazanas, Hua \& Titarchuk
(1997) to propose that the density of the hot electron cloud
responsible for the formation of the high energy spectra, through
the Comptonization process, is not uniform. Specifically they found that
the $1/f$ dependence of the hard X-ray lags on Fourier frequency
could be accounted for if the density profile of the hot scattering
medium, $n(r)$, is of the form $n(r) \propto 1/r$ for radial distance
$r$ ranging from $\sim 10^6$ to $10^{10}$ cm. Because this model keeps 
Comptonization as the mechanism for the production of high energy 
photons, it can at the same time explain both energy spectra and 
the time variablity (Hua, Kazanas \& Cui 1997). According to this
interpretation, the aperiodic variability of these sources, to which 
little attention has been paid so far, can provide diagnostics
to the density structure of the accreting gas around the black holes
(Hua, Kazanas \& Titarchuk 1997), an important information for
understanding the accretion precess (e.g. Narayan \& Yi 1994).

\end{document}